\documentclass[%
 reprint,showkeys,showpacs
 ]{revtex4-1}
\usepackage{graphicx}
\usepackage{amssymb}
\usepackage{amsmath}   %RevTexstyle

\begin{document}

%\title[Chemical Property of Colliding Sources in $^{124, 136}$Xe and $^{112, 124}$Sn Induced Collisions]
%{Chemical Property of Colliding Sources in $^{124, 136}$Xe and $^{112, 124}$Sn Induced Collisions in Isobaric Ratio Difference and Isoscaling Methods
%\footnote{Tel: +86~158 3618 1225~~Fax: +86 373 332 6691}}
\title{Chemical Property of Colliding Sources in $^{124, 136}$Xe and $^{112, 124}$Sn Induced Collisions in Isobaric Ratio Difference and Isoscaling Methods}% Force line breaks with \\

\author{Chun-Wang MA, Shan-Shan WANG, Yan-Li ZHANG, and Hui-Ling WEI}
\address{Institute of Particle and Nuclear Physics, Henan Normal University, Xinxiang, 453007 China}
\thanks{Email: machunwang@126.com}
\begin{abstract}
The isoscaling and isobaric ratio difference (IBD) methods are used to study the
$\Delta\mu/T$ ($\Delta\mu$ being the difference between the chemical potentials of neutron
and proton, and $T$ being the temperature) in the measured
1$A$ GeV $^{124}$Sn + $^{124}$Sn, $^{112}$Sn + $^{112}$Sn, $^{136}$Xe + Pb and $^{124}$Xe + Pb
reactions. The isoscaling phenomena in the $^{124}$Sn/$^{112}$Sn and the $^{136}$Xe/$^{124}$Xe
reactions pairs are investigated, and the isoscaling parameter $\alpha$ and $\beta$ are obtained. The $\Delta\mu/T$
determined by the isoscaling method (IS--$\Delta\mu/T$) and IBD method (IB--$\Delta\mu/T$) in
the measured Sn and Xe reactions are compared. It is shown that in most of fragments, the IS-- and IB-- $\Delta\mu/T$
are consistent in the Xe reactions, while the IS-- and IB-- $\Delta\mu/T$ are only similar in the less
neutron-rich fragments in the Sn reactions. The shell effects in IB--$\Delta\mu/T$ are also
discussed.

\end{abstract}

%\vspace{2pc} \noindent{\it Keywords}: isobaric yield ratio, isoscaling, symmetry energy \\
%\submitto{\JPG}
\maketitle

\section{introduction}

Research on nuclear symmetry energy has attracted much attention both theoretically
and experimentally because of its importance in nuclear physics and astrophysics (for
recent reviews, see Refs. \cite{BALi08PR,ChLWFront07,NatoPRLsym}). Among the various methods,
the yield of light particles and heavy fragments (or ratios of these
particles and fragments), are frequently used to study the symmetry energy of nuclear
matter from sub-saturation to supra-saturation densities produced in heavy-ion collisions (HICs)  \cite{ygMaSE,PuJ13PRC,Toro08IJMPE,Isoscaling,HShanPRL,Tsang-isodif-PRL04,IS-fluctuation13,Gau-flow-PRC2011,Huang10,MA12CPL09IYRAsbsbv,MaCW11PRC06IYR,MaCW12EPJA,MaCW12CPL06,MaCW13CPC}.
The results obtained from these methods constrain the nuclear symmetry energy in a
relative large scope \cite{LWEsymScope}. As the density dependence of nuclear symmetry energy is still
an open problem, an international collaboration called the "Symmetry Energy Project" has been launched in order to
determine the density dependence of the symmetry energy beyond the normal nuclear
matter based on the newly built radioactive ion beam facilities \cite{SEP}. At the
same time, it is also hoped that new new observable for studying the nuclear symmetry
energy will be found.

The chemical potentials of the neutron ($\mu_n$) and proton ($\mu_p$) depend on nuclear
density and temperature, and reflect the symmetry energy of different nuclear
matter \cite{Huang-PRC11-freeenergy,PMar12PRCIsob-sym-isos}. In HICs, the different
$\mu_n$ and $\mu_p$ in different reaction systems influence the yield of fragments,
and result in isospin phenomena in reactions \cite{FangPRC00Isospin,Luk09PRCIsospin,MaCW09PRC,MaCW09CPB}.
In contrast, $\Delta\mu\equiv(\mu_n-\mu_p)$ can be determined from
the yield of fragments. For examples, in the isoscaling and isobaric yield
ratio (IYR) methods, $\Delta\mu/T$ ($T$ being the temperature) is considered
as a probe to study the symmetry energy of nuclear matters in HICs \cite{HShanPRL,PMar12PRCIsob-sym-isos,MBTsPRL01iso,PMar-IYR-sym13PRC,MaCW13isoSB,MaCW12PRCT,Botv02PRCiso}.
As discussed in previous works, $\Delta\mu/T$ can reflect the symmetry energy
of both the colliding sources and the fragments using different scaling parameters.  \cite{Isoscaling,PMar-IYR-sym13PRC,Das05PR,Soul03-iso,Soul06-iso-T-sym,ZhouPei11T,ChenZQ10-iso-sym,Onoisoscaling,SouzaPRC09isot,Huang10NPA-Mscaling,Fang07-iso-JPG,FuY09isoCaNi}. For convenience, $\Delta\mu/T$ is also called as the "symmetry energy" in this
work. The $\Delta\mu/T$ determined by the isoscaling (IS--$\Delta\mu/T$) and the
isobaric yield ratio difference (IBD) methods (IB--$\Delta\mu/T$) have been
compared in previous works \cite{PMar12PRCIsob-sym-isos,PMar-IYR-sym13PRC,MaCW13isoSB,Mallik13-sym-IYR}.
Similar distributions of the IB--$\Delta\mu/T$ and IS--$\Delta\mu/T$ are found,
which also reveals that $\Delta\mu/T$ depends on the neutron density ($\rho_n$)
and the proton density ($\rho_p$) distributions, as well as the asymmetry of projectile.

In comparison with the fragments produced by intermediate projectile fragmentation in
Ref. \cite{MaCW13isoSB}, which have relative small masses, we will analyze the fragments
produced by the $^{124, 136}$Xe and $^{112, 124}$Sn projectile fragmentation,
which have larger masses and neutron-excess. This paper is organized as follows.
In Sec. \ref{model}, the isobaric and isoscaling methods to determine $\Delta\mu/T$
are introduced; in Sec. \ref{result}, $\Delta\mu/T$ obtained from the studies
of fragments in the $^{124, 136}$Xe and $^{112, 124}$Sn reactions are reported
and discussed; In Sec. \ref{summary}, a summary is presented.

\section{model description}
\label{model}

The IBD method has previously been described in Ref. \cite{MaCW13isoSB}. To illustrate
the results more clearly, we briefly introduce the isoscaling and IBD methods based on
the grand-canonical ensembles theory. The yield of a fragment with mass $A$ and neutron-excess
$I$ ($I = N-Z$) in the grand-canonical limit is given by \cite{GrandCan,Tsang07BET},
\begin{equation}\label{yieldGC}
Y(A,I) = C{_0}A^{\tau}exp\{[-F(A,I,T)+\mu_{n}N+\mu_{p}Z]/T\},
\end{equation}
where $C{_0}$ is a constant; $\mu_n$ and $\mu_p$ are the chemical potential of neutron
and proton, and $N$ and $Z$ are the neutron and proton numbers. $\tau$ is nonuniform in
different reaction systems \cite{Huang10Powerlaw}. $F(A,I,T)$ is the free energy of
the cluster (fragment), and $T$ is the temperature.

The isoscaling phenomena are shown in the isotopic and isotonic ratios between two
reactions of similar measurements. From Eq. (\ref{yieldGC}), the yield ratio of one
fragment between two reactions, $R_{21}^{IS}(N,Z)$ (1 and 2 denoting the reactions,
usually 2 denoting the neutron-rich reaction system), can be defined as,
\begin{equation}\label{IS1}
R_{21}^{IS}(N,Z)=Y_{2}(N,Z)/Y_{1}(N,Z)=C'\mbox{exp}(\alpha N+ \beta Z),
\end{equation}
Taking the logarithm of Eq. (\ref{IS1}), one can obtain,
\begin{equation}\label{IS2}
\mbox{ln}R_{21}^{IS}(N,Z)=C(\alpha N+ \beta Z),
\end{equation}
where $C'$ and $C$ are overall normalization constants which originate from the property
of reaction system. $\mu_n$ and $\mu_p$ are assumed to change very slowly, and can thus be seen
as the same in one reaction; $\alpha = \Delta\mu_n/T$ with $\Delta \mu_n = \mu_{n2} - \mu_{n1}$,
and $\beta = \Delta\mu_p/T$ with $\Delta\mu_p = \mu_{p2} - \mu_{p1}$, which reflect the
properties of colliding source. In the isotopic ratio, $\beta$ cancels out and $\alpha$ can
be fitted; and in the isotonic ratio, $\alpha$ cancels out and $\beta$ can be fitted. $\alpha$
can be related to the symmetry energy ($C_{sym}$) in nuclear mass of colliding source by
different scaling parameters \cite{IS-fluctuation13,PMar-IYR-sym13PRC,ChenZQ10-iso-sym,Onoisoscaling,SouzaPRC09isot,Huang10NPA-Mscaling,BA06IS-rev}.

\begin{figure*}
\begin{center}
\includegraphics[width=14cm]{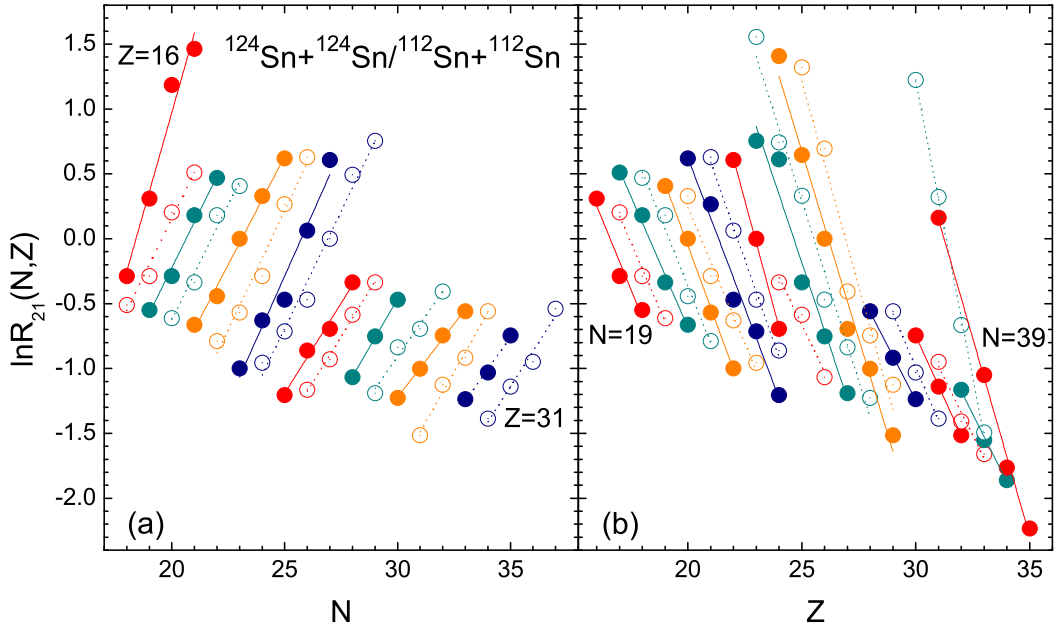}
\caption{(Color online)\label{Snscaling}
Isoscaling phenomena between fragments in the 1$A$ GeV $^{124}$Sn + $^{124}$Sn
and $^{112}$Sn + $^{112}$Sn reactions \cite{Fhor11Sndata}. From left to right,
in (a) the plotted are isotopic ratio of isotopes with $Z$ increase from to 16
to 31, and (b) isotonic ratio of isotones with $N$ from 19 to 39. The linear
fits to the isoscalings are shown as lines.
}
\end{center}
\end{figure*}

For the IBD method, first, the IYR between isobars differing by 2 units in $I$, $R^{IB}(I+2,I,A)$, can be defined in single reaction as,
\begin{eqnarray}\label{ratiodef}
&R^{IB}(I+2,I,A)
=Y(A,I+2)/Y(A,I)  \nonumber\\
& =\mbox{exp}\{[F(I+2,A,T)-F(I,A,T)+\mu_n-\mu_p]/T\},
\end{eqnarray}
Unlike in the isoscaling method, the $C_{0}A{^\tau}$ term in Eq. (\ref{yieldGC}) cancels
out and the system dependence is removed in single reaction. Assuming that the isobars involved
in the ratio have the same temperature, only the retained $\mu_n$ and $\mu_p$ are related to the
colliding sources. Taking the logarithm of Eq. (\ref{ratiodef}), one obtains,
\begin{equation}\label{tlnRcal}
\mbox{ln}R^{IB}(I+2,I,A)=(\Delta F+\Delta\mu)/T,
\end{equation}
where $\Delta F = F(I+2,A,T)- F(I,A,T)$, and $\Delta\mu=\mu_n-\mu_p$ as previously defined. Now we
can define the difference between IYRs, which is labeled as the IBD, in two reactions as the follows,
\begin{eqnarray}\label{DBratio}
\Delta \mbox{ln}R_{21}^{IB}&=\mbox{ln}[R_{2}^{IB}(I+2,I,A)]-\mbox{ln}[R_{1}^{IB}(I+2,I,A)]   \nonumber\\
             & =[(\mu_{n2}-\mu_{n1})-(\mu_{p1}-\mu_{p2})]/T=\Delta\mu_{n}/T-\Delta\mu_{p}/T     \nonumber\\
             & =\alpha-\beta, \hspace{0.4cm}
\end{eqnarray}
Eq. (\ref{DBratio}) reveals the relationship between the $\Delta\mu/T$ determined by
isoscaling and IBD methods, i.e., IB--$\Delta\mu/T\equiv\Delta\mbox{ln}R_{21}^{IB}$ and
IS--$\Delta\mu/T\equiv\alpha-\beta$, and IB--$\Delta\mu/T$ should equal the IS--$\Delta\mu/T$.
%
%\begin{figure}
%\begin{center}
%\includegraphics[width=13cm]{fig1Snscaling}
%\caption{(Color online)\label{Snscaling}
%Isoscaling phenomena between fragments in the 1$A$ GeV $^{124}$Sn + $^{124}$Sn
%and $^{112}$Sn + $^{112}$Sn reactions \cite{Fhor11Sndata}. From left to right,
%in (a) the plotted are isotopic ratio of isotopes with $Z$ increase from to 16
%to 31, and (b) isotonic ratio of isotones with $N$ from 19 to 39. The linear
%fits to the isoscalings are shown as lines.
%}
%\end{center}
%\end{figure}

\begin{figure*}
\begin{center}
\includegraphics[width=14cm]{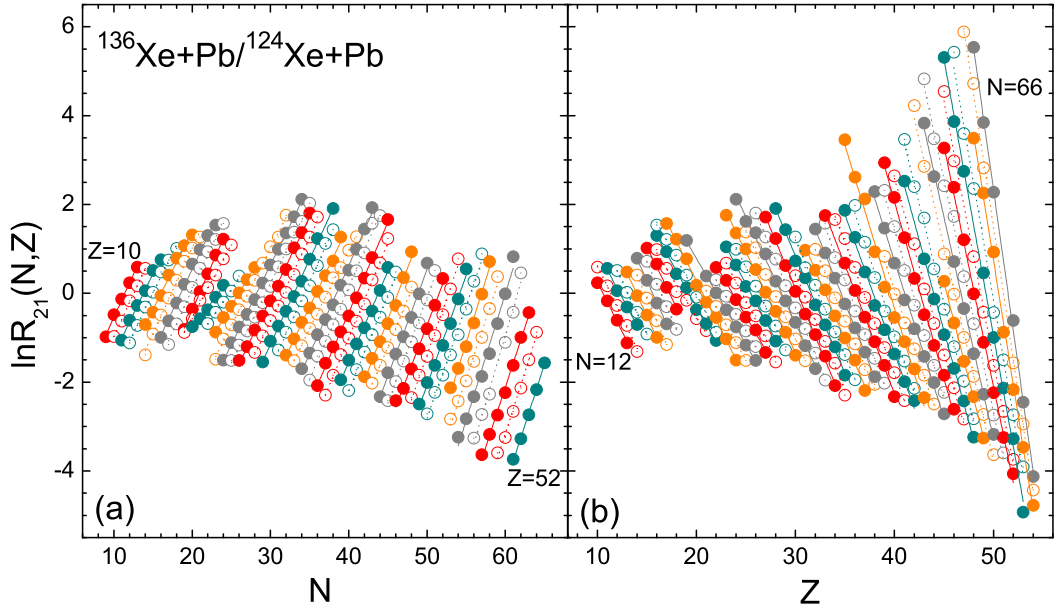}
\caption{(Color online)\label{Xescaling}
Isoscaling phenomena in the fragment ratios between the 1$A$ GeV $^{136}$Xe + Pb and $^{124}$Xe + Pb
reactions \cite{Henz08}. From left to right, in (a) the atomic numbers $Z$ increase from to 10
to 52, and in (b) the neutron numbers $N$ increase from 12 to 66. The linear fits to the
isoscalings are shown as lines.
}
\end{center}
\end{figure*}

\section{Results and discussions}
\label{result}

In this paper, the yields of fragments in the measured 1$A$ GeV $^{124}$Sn + $^{124}$Sn,
$^{112}$Sn + $^{112}$Sn reactions \cite{Fhor11Sndata}, and 1$A$ GeV $^{124, 136}$Xe + Pb
\cite{Henz08} will be analyzed using the isoscaling and IBD methods described above.
First, the isoscaling phenomena in these reactions will be illustrated. The isotopic and
isotonic ratios in the Sn and Xe reactions will also be analyzed to extract the isoscaling
parameters $\alpha$ and $\beta$ according to Eq. (\ref{IS2}).
In Fig. \ref{Snscaling}, the isotopic and isotonic ratios between the $^{124}$Sn + $^{124}$Sn
and $^{112}$Sn + $^{112}$Sn reactions are plotted, in which $Z$ changes from 16 to 31 and
$N$ changes from 19 to 39. The different isotopes are shown by alternating full and open
symbols for clarity. Quite good linear scalings are shown in the isotopic and isotonic ratios,
which are denoted as lines from the linear fitting. In Fig.\ref{Snscaling}(b), the ratios
of the isotones from $N = 33$ to 37 show different trends to those of other $N$--isotones,
at the same time, less isotones are measured than the other isotones.

In Fig. \ref{Xescaling}, the isotopic and isotonic ratios between the 1$A$ GeV $^{136}$Xe + Pb
and $^{124}$Xe + Pb reactions are plotted, in which $Z$ changes from 10 to 52 and $N$
changes from 12 to 66. Much more fragments in the Xe reactions are measured than the Sn
reactions. The ratios of the $Z = 20$ and 21 isotopes show a little different trend to
those of other Z-isotopes, at the same time, less isotopes are measured than
other $Z$--isotopes.

\begin{figure}
\begin{center}
\includegraphics[width=8cm]{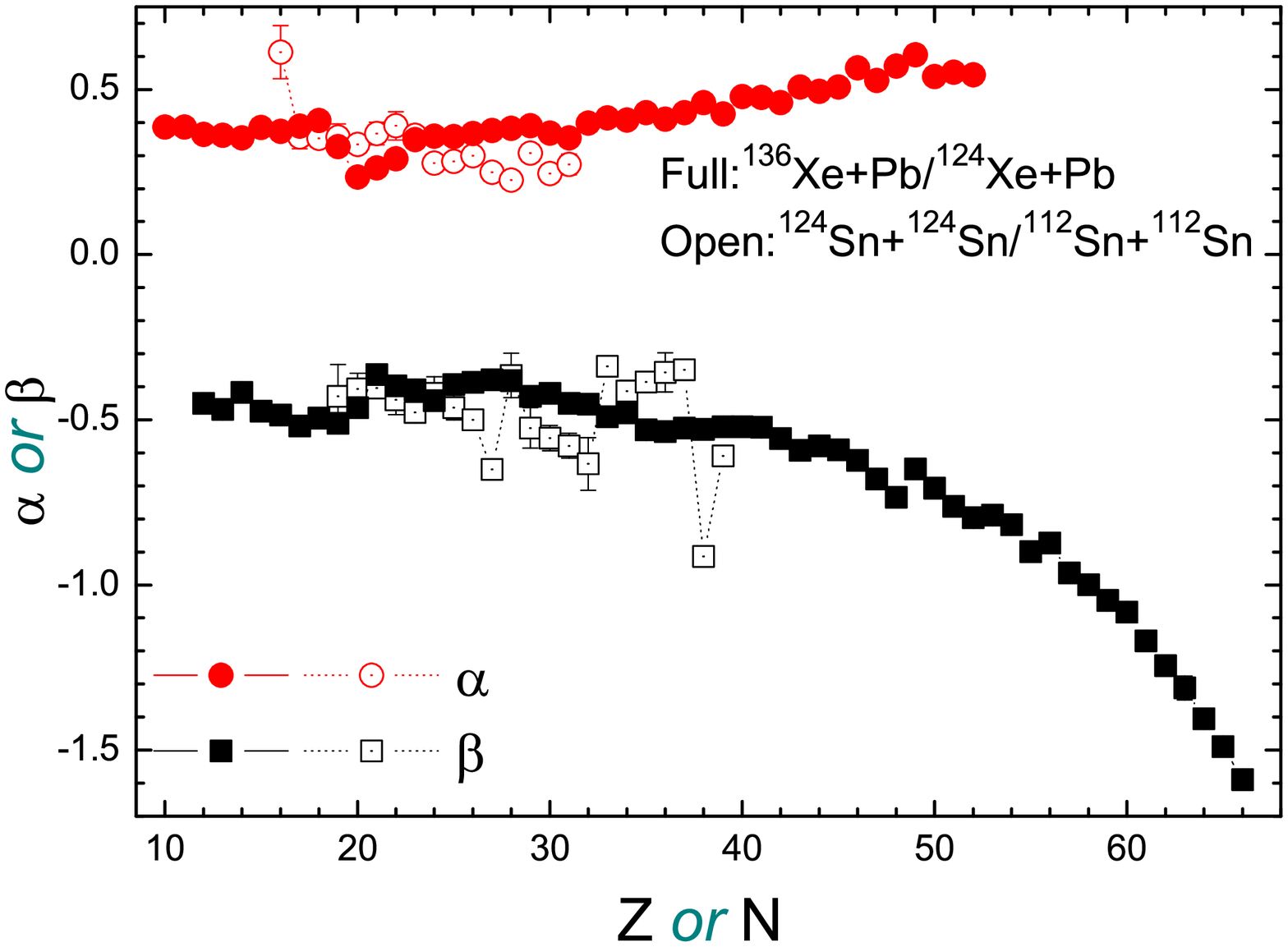}
\caption{\label{alphabeta} (Color online)
The results of the isoscaling parameters $\alpha$ and $\beta$ from the linear fitting in
Figs. \ref{Snscaling} and \ref{Xescaling}. The full and open symbols represent the results
of the Xe and Sn reactions, respectively; the circles and squares represent $\alpha$ and
$\beta$, respectively.
}
\includegraphics[width=8cm]{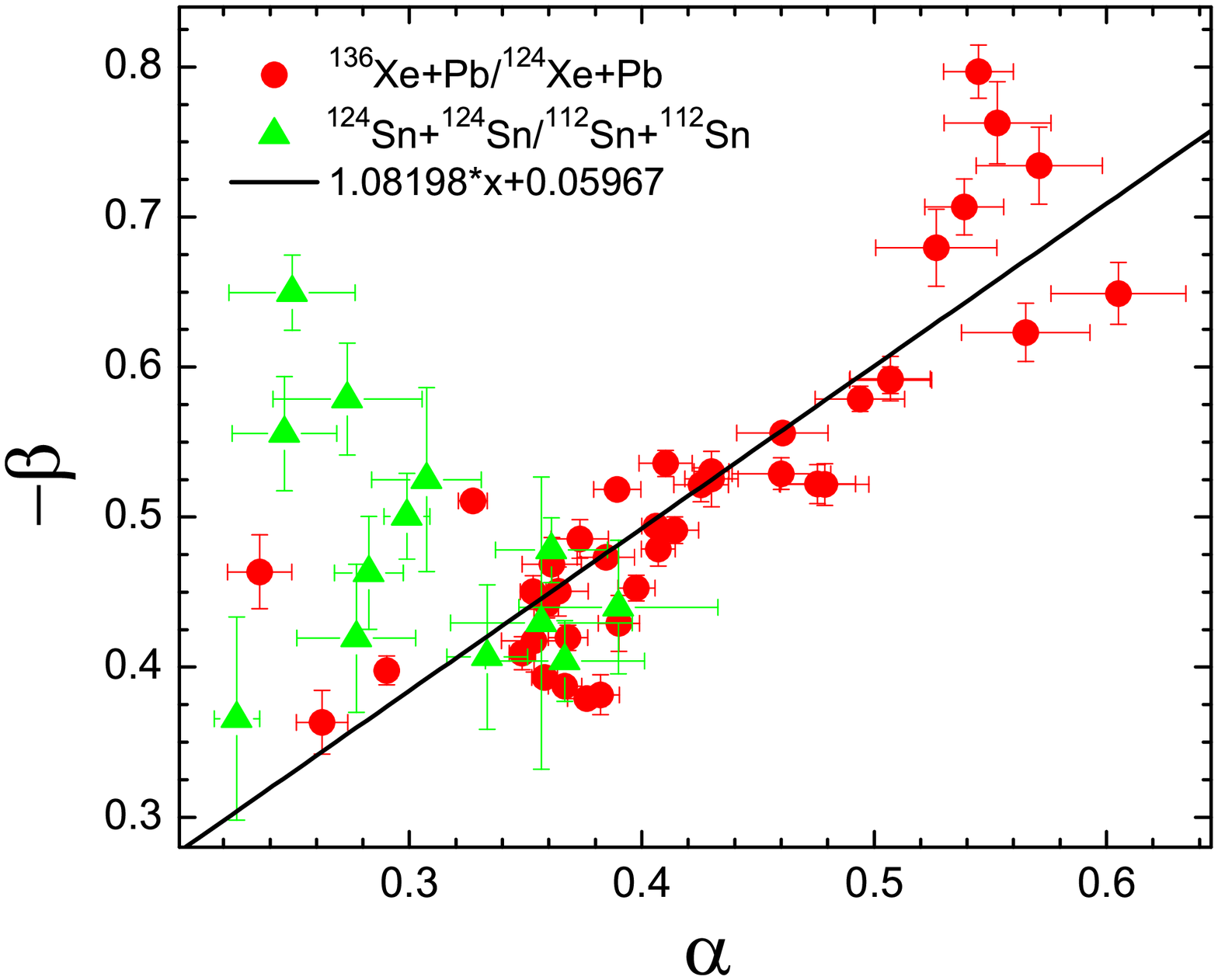}
%{pp31asym-T-inall}
\caption{\label{Corr-a-b}
(Color online) The correlations between the isoscaling parameters $\alpha$ and
$-\beta$ shown in Fig. \ref{alphabeta} from the fragments in the Sn and Xe
reactions. The line represents the linear fits to the $\alpha$ and $-\beta$
correlation in the Xe reactions.
}
\end{center}
\end{figure}

In Fig. \ref{alphabeta}, the obtained isoscaling parameters $\alpha$ and $\beta$ are plotted. The values
of $\alpha$ form a plateau around 0.4 in a wide range of $Z$ when $Z < 40$ both in the Sn and Xe reactions.
When $Z>40$, $\alpha$ increases with $Z$ in the Xe reactions. In particular, there is a minimum value
around $Z = 20$ in the Xe reactions. For isotones, $\beta$ shows similar trend except the values are negative.
When $N < 40$, $\beta$ forms a plateau around -0.45, and then decreases with increasing $N$. The $\beta$
distribution in the Sn reactions is not as smooth as that in the Xe reactions. The $\alpha$--plateau in the
25$A$ MeV $^{86}$Kr + $^{124, 112}$Sn reactions is around 0.4 \cite{Soul06-iso-T-sym}, which coincides the
result in this work. A decreasing of $\alpha$ is observed, however, when $Z > 30$, which is not shown due to the
absence of data in this work. $\alpha = 0.36$ and $\beta = -0.4$ are obtained in the measured central
$^{112}$Sn + $^{112}$Sn and $^{124}$Sn + $^{124}$Sn collisions at an incident energy of 50$A$ MeV \cite{Das05PR}.
The $\alpha$ and $\beta$ values are also coincident with the results of the Sn reactions in this work.

Generally, $-\beta \approx \alpha$ is assumed \cite{PMar12PRCIsob-sym-isos,FuY09isoCaNi}. For one
fragment, specific $\alpha$ and $\beta$ can be obtained from its $Z$-isotopes and $N$-isotones,
respectively, and the correlation between $\alpha$ and $\beta$ can be obtained. The correlation
between $-\beta$ and $\alpha$ in Sn and Xe reactions is plotted in Fig. \ref{Corr-a-b}. The
$-\beta$ and $\alpha$ correlation in Xe reactions is fitted using a linear function, which
reads $y = 1.08x + 0.06$ and satisfies the approximation $-\beta \approx \alpha$. For the Sn reactions,
$-\beta$ does not depend on $\alpha$ linearly, thus the correlation is not fitted.

\begin{figure*}
\begin{center}
\includegraphics[width=15cm]{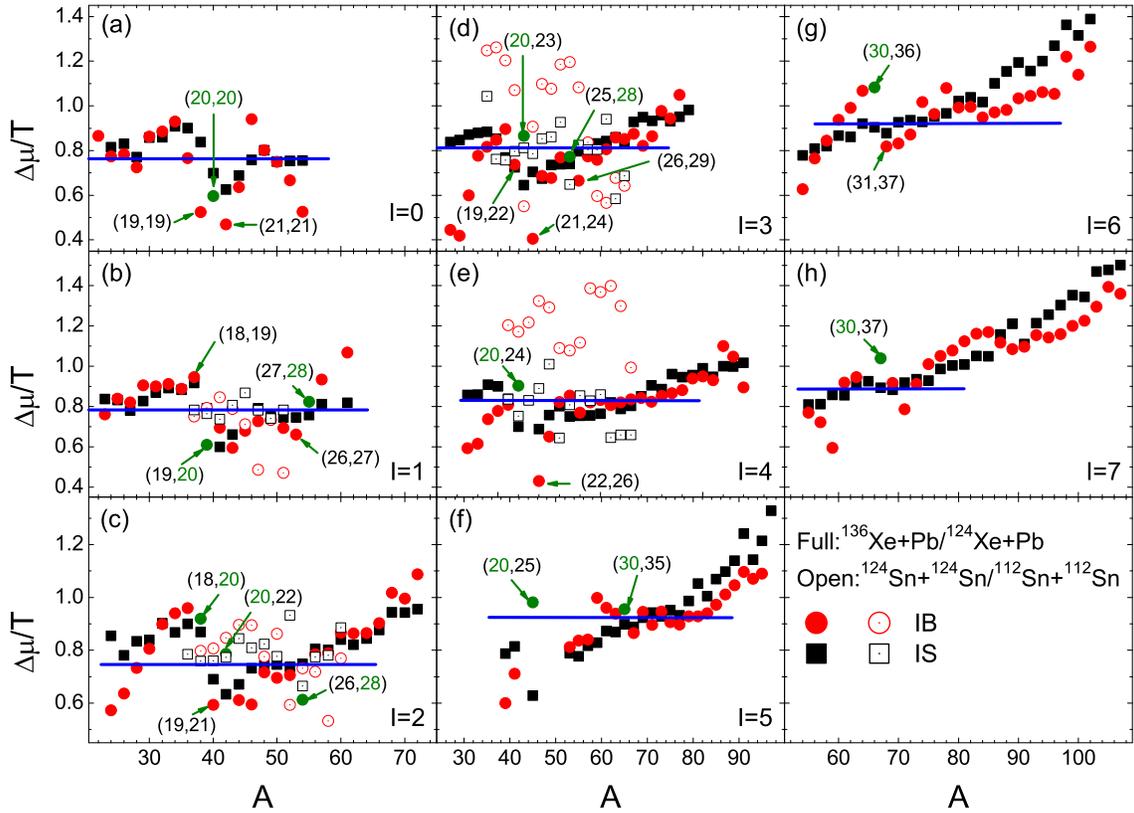}\caption{(Color online)
The values of IS-- and IB-- $\Delta\mu/T$ determined from the measured
fragments ratios in Sn and Xe reactions. The numbers in brackets label
the protons and neutrons of the nucleus $(Z, N)$ (see the text for explanation).
The green circles represent the fragments having a magic number. The lines
are just for guiding the eyes.
}\label{DmuT}
\end{center}
\end{figure*}

Second, the IS-- and IB-- $\Delta\mu/T$ results will be compared. From Eq. (\ref{DBratio}),
IS--$\Delta\mu/T$ should equal to $\alpha-\beta$. The value of IB--$\Delta\mu/T$ is calculated
from IYRs according to Eq. (\ref{DBratio}). The values of $\Delta\mu/T$ are plotted in
Fig. \ref{DmuT}. Trends of $\Delta\mu/T$ distributions similar to the Ca and Ni reactions
are found \cite{MaCW13isoSB}. Generally, $\Delta\mu/T$ also shows the distribution of
a plateau plus an increasing part. In most of the fragments, IB--$\Delta\mu/T$ agrees well with
IS--$\Delta\mu/T$ in the Xe reactions. In the Sn reactions, IB--$\Delta\mu/T$ only
agrees with IS--$\Delta\mu/T$ in the $I = $1 and 2 fragments, but a relative large difference
between them are shown in the $I = $3 and 4 fragments. The plateaus form around $\Delta\mu/T = 0.8$,
as shown by the guiding lines. The $\Delta\mu/T$ plateaus in the $^{48}$Ca/$^{40}$Ca, $^{64}$Ni/$^{58}$Ni,
$^{58}$Ni/$^{40}$Ca, and  $^{48}$Ca/$^{64}$Ni reactions are around 2.0, 1.4, 0.7, and 0.5 \cite{MaCW13isoSB}.
The height of the plateau is explained as being due to the difference between the $\rho_n$ distributions, as well
as the differences between the $\rho_p$ distributions in the core of the projectiles. The $\Delta\mu/T$
in the Sn and Xe reactions are close to that of the $^{58}$Ni/$^{40}$Ca reactions, which indicates
that in the core of $^{124}$Sn/$^{112}$Sn and $^{136}$Xe/$^{124}$Xe, the difference between the $\rho_n$
and $\rho_p$ distributions is similar to those between the cores of $^{58}$Ni/$^{40}$Ca.

Finally, we discuss the possible shell effects in IB--$\Delta\mu/T$ shown
in Fig. \ref{DmuT}, which are not obvious in the Ca and Ni reactions \cite{MaCW13isoSB}.
In a previous study, the shell effect is also found in the symmetry-energy coefficient
($a_{sym}$) of nucleus \cite{MA12CPL09IYRAsbsbv}, which results in sudden changes in $a_{sym}$.
Generally, $\mu_n$, $\mu_p$ and the resultant $\Delta\mu$ should vary very little, and
there should be no nuclear structure effects in the colliding process near the critical
point temperature \cite{Huang-PRC11-freeenergy}. In actual reactions, however, the primary
fragment are greatly influenced by the secondary decay, and the nuclear structure effects
can be manifested after the decay process. A relative low temperature ($T\sim2 MeV$) is
found for fragments with large mass using an isobaric ratio method \cite{MaCW12PRCT}.
At this low temperature, the shell effects may be obvious in $\Delta\mu/T$. It is clearly
observed that around the magic number $Z = 20$, the isotopic ratios in the Xe reactions
show different trends to the other isotopes. The same also happens in the $N = 20$ and
$\sim 30$ isotonic ratios, which makes $\alpha$ and $\beta$ change suddenly, thus result
in large gaps in $\Delta\mu/T$ of fragments with a magic number. IB--$\Delta\mu/T$,
as has been noted, is more sensitive to the variation of symmetry energy than IS--$\Delta\mu/T$
since the IBD method uses only two isobars \cite{MaCW13isoSB}. Taking IB--$\Delta\mu/T$
in the Xe reactions as an example, most of the IB--$\Delta\mu/T$ can be divided into different
bands according to the shells. Some fragments with a magic number (or near the magic number)
are labeled as ($Z, N$) in Fig. \ref{DmuT}. Large gaps are shown between the IB--$\Delta\mu/T$
near the closed-shell fragment and its neighbor, for example, the $I=1$ fragments (18, 19)
and (19, \textbf{20}) in (b); the $I=2$ fragments (18, \textbf{20}) and (19, 21) in (c);
the $I=3$ fragments (25, \textbf{28}) and (26, 29) in (d). The shell effects of $N = 28$
are not obvious when $I \geq 3$, and the same happens in $Z = 28$ fragment. The violent
behavior is also observed in IB--$\Delta\mu/T$ of the neutron-rich $I = $6 and $I = $7 fragments
within the range of $Z = 26 \sim30$.

\section{Summary}
\label{summary}

In this paper, the isoscaling and the IBD methods are used to study the $\Delta\mu/T$ by analyzing the
fragment yield in the measured 1$A$ GeV $^{124}$Sn + $^{124}$Sn, $^{112}$Sn + $^{112}$Sn, $^{136}$Xe + Pb and
$^{124}$Xe + Pb reactions. First, the isoscaling phenomena in the $^{124}$Sn/$^{112}$Sn and
$^{136}$Xe/$^{124}$Xe reactions are analyzed, and the isoscaling parameter $\alpha$ and $\beta$
are obtained. $\alpha$ ($\beta$) is found to be almost constant in Sn reactions. The $\alpha$
($\beta$) also changes very little in Xe reactions when $Z < 40$ ($N < 40$), but $\alpha$ ($\beta$)
increases (decreases) when $Z > 40$ ($N > 40$). The $\alpha \approx -\beta$ approximation is satisfied
in Xe reactions but not satisfied in Sn reactions.

The results of IS--$\Delta\mu/T$ and IB--$\Delta\mu/T$ in Sn and Xe reactions are compared.
In most of the fragments, the IS-- and IB-- $\Delta\mu/T$ are consistent in Xe reactions, while they
are only similar in the $I =$ 1 and 2 fragments in Sn reactions. The values of the $\Delta\mu/T$ plateaus
indicate that the difference between $\rho_n$ and $\rho_p$ distributions in the core of $^{136}$Xe/$^{124}$Xe
and $^{124}$Sn/$^{112}$Sn is similar as that of the $^{58}$Ni/$^{40}$Ca.

The possible shell effects in the IB--$\Delta\mu/T$ of fragments are also discussed. The IB--$\Delta\mu/T$ in the Xe reactions
can be divided into different bands according to different shells. The $Z =$ 20 and $N =$28 shell effects
are obvious in IB--$\Delta\mu/T$ when the fragments are not very neutron-rich.

\section*{Acknowledgements}
This work is supported by the National Natural Science Foundation of China (Grant No. 10905017),
Program for Science \& Technology Innovation Talents in Universities of Henan Province (13HASTIT046),
and the Young Teacher Project in Henan Normal University.

%\section*{References}


\begin{thebibliography}{46}

\bibitem{BALi08PR}
Li B A, Chen L W, Ko C M, 2008
%{\it et al.},
Phys. Rep. \textbf{464} 113.

\bibitem{ChLWFront07} Chen L W \textit{et al.}, 2007 Front. Phys. China \textbf{2} 327.
\bibitem{NatoPRLsym}
Natowitz J B, R\"{o}pke G, Typel S %,
%D. Blaschke, A. Bonasera, K. Hagel, T. Kl\"{a}hn, S. Kowalski, L. Qin, S. Shlomo, R. Wada, and H. H. Wolter
{\it et al.}, 2010
Phys. Rev. Lett. \textbf{104} 202501.
\bibitem{ygMaSE} Kumar S, Ma Y G, Zhang G Q \textit{et al}, 2012 %, C. L. Zhou, Phys.
Rev. C \textbf{85} 024620; 2011 \textit{ibid,} \textbf{84} 044620; 2012 \textit{ibid,} \textbf{86} 051601(R).
%---------------double ratio(n/p)----------------------
\bibitem{PuJ13PRC}
Pu J, Chen J H, Kumar S %, Y. G. Ma, C. W. Ma and G. Q. Zhang,
\textit{et al},
2013 Phys. Rev. C \textbf{87} 047603.

\bibitem{Toro08IJMPE} Toro M D \textit{et al}, 2008 Int. J. Mod. Phys. E \textbf{17} 1799.

\bibitem{Isoscaling}
Ma Y G, Wang K, Wei Y B %,
%G. L. Ma, X. Z. Cai, J. G. Chen, D. Q Fang, W. Guo, W. Q. Shen, W. D. Tian, and C. Zhong
\textit{et al.},
2004 Phys. Rev. C \textbf{69} 064610.

\bibitem{HShanPRL}
Xu H S,  Tsang M B, Liu T X %, X. D. Liu, W. G. Lynch, W. P. Tan, A. Vander Molen, G. Verde, A. Wagner, H. F. Xi, C. K. Gelbke,
%L. Beaulieu, B. Davin, Y. Larochelle, T. Lefort, R. T. de Souza, R. Yanez, V. E. Viola, R. J. Charity and L. G. Sobotka,
2000 \textit{et al.}, Phys. Rev. Lett. \textbf{85} 716.

\bibitem{Tsang-isodif-PRL04} Tsang M B \textit{et al.}, 2004 Phys. Rev. Lett. \textbf{92} 062701;
Famiano M A \textit{et al.}, 2006 Phys. Rev. Lett. \textbf{97} 052701;
Tsang M B \textit{et al.}, 2009 Phys. Rev. Lett. \textbf{102} 122701.

\bibitem{IS-fluctuation13} Colonna M, 2013 Phys. Rev. Lett. \textbf{110} 042701.

\bibitem{Gau-flow-PRC2011}
Gautam S, Sood A D, Puri R K, Aichelin  J
% \textit{et al.},
2011 Phys. Rev. C \textbf{83} 034606.

%----------------------IYR------------------
\bibitem{Huang10}
Huang M, Chen Z, Kowalski S %,
%Y. G. Ma, R. Wada, T. Keutgen,
%K. Hagel, M. Barbui, A. Bonasera, C. Bottosso, T. Materna, J. B. Natowitz,
%L. Qin, M. R. D. Rodrigues, P. K. Sahu, and J. Wang,
{\it et al.},
2010 Phys. Rev. C \textbf{81} 044620.

\bibitem{MA12CPL09IYRAsbsbv}
Ma C W, Yang J B, Yu M %, PU
%Jie, WANG Shan-Shan, and WEI Hui-Ling
{\it et al.},
2012 Chin. Phys. Lett. \textbf{29} 092101.

\bibitem{MaCW11PRC06IYR}
Ma C W, Wang F, Ma Y G and Jin C,
 % {\it et al.},
2011 Phys. Rev. C \textbf{83} 064620.

\bibitem{MaCW12EPJA}
Ma C W, Pu J, Wei H L %, Shan-Shan Wang, Heng-Li Song,
%Sha Zhang, and Li Chen,
{\it et al.},
2012 Eur. Phys. J. A \textbf{48} 78. % doi:10.1140/epja/i2012-12078-5

\bibitem{MaCW12CPL06}
Ma C W, Pu J, Wang S S, and Wei H L
 % {\it et al.},
2012 Chin. Phys. Lett. \textbf{29} 062101.

\bibitem{MaCW13CPC}
Ma C W, Song H L, Pu J %, T. L. ZHANG, S. ZHANG, S. S. WANG, X. L. ZHAO, L. CHEN,
{\it et al.},
2013 Chin. Phys. C \textbf{37} 024102.


\bibitem{LWEsymScope} Chen L W, arXiv:1212.0284 [nucl-th].
\bibitem{SEP} https://groups.nscl.msu.edu/hira/sep.htm

\bibitem{Huang-PRC11-freeenergy}
Huang M, Bonasera A, Chen Z %, R. Wada, K. Hagel, J. B. Natowitz,
%P. K. Sahu, L. Qin, T. Keutgen, S. Kowalski, T. Materna,
%J. Wang, M. Barbui, C. Bottosso, and M. R. D. Rodrigues,
{\it et al.},
2011 Phys. Rev. C \textbf{81} 044618.

\bibitem{PMar12PRCIsob-sym-isos} % IYR, isoscaling, m-scaling, \Delta-in-scaling
Marini P, Bonasera A, McIntosh A %, R. Tripathi, S. Galanopoulos,
%K. Hagel, L. Heilborn, Z. Kohley, L. W. May, M. Mehlman, S. N. Soisson,
%G. A. Souliotis, D. V. Shetty, W. B. Smith, B. C. Stein, S. Wuenschel,
%and S. J. Yennello
{\it et al.},
2012 Phys. Rev. C \textbf{85} 034617.

\bibitem{FangPRC00Isospin}
Fang D Q, Shen W Q, Feng J \textit{et al.},
2000 Phys. Rev. C \textbf{61} 044610.
\bibitem{Luk09PRCIsospin}
Lukyanov S, Mocko M, Andronenko L
\textit{et al.},
2009 Phys. Rev. C \textbf{80} 014609.
\bibitem{MaCW09PRC}
Ma C W, Wei H L, Wang J Y %,
%G. J. Liu, Y. Fu, D. Q. Fang, W. D. Tian, X. Z. Cai, H. W. Wang, and Y. G. Ma,
{\it et al.},
2009 Phys. Rev. C \textbf{79} 034606.
\bibitem{MaCW09CPB}
Ma C W, Wei H L, and Wang J Y,
 %{\it et al.},
2009 Chin. Phys. B \textbf{18} 4781.


\bibitem{MBTsPRL01iso}
Tsang M B, Friedman W A, Gelbke C K %, W. G. Lynch,
%G. Verde, and H. Xu,
\textit{et al.},
2001 Phys. Rev. Lett. \textbf{86} 5023.

\bibitem{PMar-IYR-sym13PRC} %IYR -isoscaling- multifragmentation-smm-gemini
Marini P, Bonasera A, Souliotis G A %, P. Cammarata, S. Wuenschel,
%R. Tripathi, Z. Kohley, K. Hagel, L. Heilborn, J. Mabiala, L. W. May,
%A. B. McIntosh, and S. J. Yennello
{\it et al.},
2013 Phys. Rev. C \textbf{87} 024603.

\bibitem{MaCW13isoSB} %isoscaling-IYR-similarity \Delta_{np}/T
Ma C W, Wang S S, Zhang Y L, Wei H L,
2013 Phys. Rev. C \textbf{87} 034618.

\bibitem{MaCW12PRCT}
Ma C W, Pu J, Ma Y G %, R. Wada, and S. S. Wang,
{\it et al.},
2012 Phys. Rev. C \textbf{86} 054611;
Ma C W, Zhao X L, Pu J %, Wang S S, Qiao C Y, Feng X, Wada R, Ma Y G,
{\it et al.},
2013 Phys. Rev. C \textbf{88} 014609.

\bibitem{Botv02PRCiso}
Botvina A S, Lozhkin O V, and Trautmann W %,
\textit{et al.},
2002 Phys. Rev. C \textbf{65} 044610.

\bibitem{Das05PR}
Das C B, Gupta S D, Lynch W G %, A.Z. Mekjian, and M.B. Tsang,
{\it et al.},
2005 Phys. Rep. \textbf{406} 1.

\bibitem{Soul03-iso} %isoscaling, 25AMeV  Sn reactions
Souliotis G A, Shetty D V, Veselsky M %,
%G. Chubarian, L. Trache, A. Keksis, E. Martin, and S. J. Yennello,
\textit{et al.},
2003 Phys. Rev. C \textbf{68} 024605.
\bibitem{Soul06-iso-T-sym}
Souliotis G A, Shetty D V, Keksis A %, E. Bell, M. Jandel, M. Veselsky, S.J. Yennello,
\textit{et al.},
2006 Phys. Rev. C \textbf{73} 024606.
\bibitem{ZhouPei11T}
P Zhou, W D Tian, Y G Ma %,X. Z. Cai, D. Q. Fang, and H. W. Wang
{\it et al.},
2011 Phys. Rev. C \textbf{84} 037605.
\bibitem{ChenZQ10-iso-sym}
Chen Z, Kowalski S, Huang M %, R. Wada,
%T. Keutgen, K. Hagel, A. Bonasera, J. B. Natowitz, T. Materna,
%L. Qin, P. K. Sahu, and J. Wang,
\textit{et al.},
2010 Phys. Rev. C \textbf{81} 064613.
\bibitem{Onoisoscaling}
A. Ono, P. Danielewicz, W. A. Friedman %, W. G. Lynch, and M. B. Tsang,
{\it et al.}, 2003 Phys. Rev. C \textbf{68} 051601(R); 2004  \textit{ibid}, \textbf{70} 041604(R).

\bibitem{SouzaPRC09isot} %temperature effect in isoscaling
Souza S R, Tsang M B, Carlson  B V %, R. Donangelo, W. G. Lynch, and A. W. Steiner
\textit{et al.},
2009 Phys. Rev. C \textbf{80} 044606.

\bibitem{Huang10NPA-Mscaling}
Huang M, Chen Z, Kowalski S %, R. Wada, T. Keutgen,
%K. Hagel, J. Wang, L. Qin, J.B. Natowitz, T. Materna,
%P.K. Sahu, M. Barbui, C. Bottosso, M.R.D. Rodrigues, A. Bonaser
{\it et al.},
2011 Nucl. Phys. A \textbf{847} 233. %233-242

\bibitem{BA06IS-rev} Li B A, and Chen L W 2006 Phys. Rev. C \textbf{74} 034610.

\bibitem{Fang07-iso-JPG}
Fang D Q, Ma Y G, Zhong C %, C. W. Ma,
%X. Z. Cai, J. G. Chen, W. Guo, Q. M. Su, W. D. Tian, K. Wang, T. Z. Yan and W. Q. Shen
{\it et al.},
2007 J. Phys. G: Nucl. Part. Phys. \textbf{34} 2173.
\bibitem{FuY09isoCaNi}%isoscaling
Fu Y, Fang D Q, Ma Y G%, Cai Xiang-Zhou, Tian Wen-Dong, Wang Hong-Wei, and Guo Wei
{\it et al.},
2009 Chin. Phys. Lett. \textbf{26} 082503.

\bibitem{Mallik13-sym-IYR}
Mallik S and Chaudhuri G,
2013 Phys. Rev. C \textbf{87} 011602(R).

%%-----------------fragment formation temperature esembles theory-----------
\bibitem{GrandCan} %Comparison of canonical and grand canonical models
Das C B, Gupta S D, Liu X D, and Tsang M B
 %{\it et al.},
2001 Phys. Rev. C \textbf{64} 044608.

\bibitem{Tsang07BET} %T=2.2MeV Canonical thermodynamic model
Tsang M B, Lynch W G, Friedman W A %,
%M. Mocko, Z. Y. Sun, N. Aoi, J. M. Cook, F. Delaunay,
%M. A. Famiano, H. Hui, N. Imai, H. Iwasaki, T. Motobayashi,
%M. Niikura, T. Onishi, A. M. Rogers, H. Sakurai, H. Suzuki,
%E. Takeshita, S. Takeuchi, and M. S. Wallace,
\textit{et al.},
2007 Phys. Rev. C \textbf{76} 041302(R).
\bibitem{Huang10Powerlaw} %tau determination
Huang M, Wada R, Chen Z %, T. Keutgen, S. Kowalski, K. Hagel,
%M. Barbui, A. Bonasera, C. Bottosso, T. Materna, J. B. Natowitz,
%L. Qin, M. R. D. Rodrigues, P. K. Sahu, K. J. Schmidt, and J. Wang
\textit{et al.},
2010 Phys. Rev. C \textbf{82} 054602(R).
%---------------experimental data---------------------

\bibitem{Fhor11Sndata}
F\"{o}hr V, Bacquias A, Casarejos E %, T. Enqvist, A. R. Junghans,
%A. Keli\'{c}-Heil, T. Kurtukian, S. Luki\'{c},D. P\'{e}rez-Loureiro,
%R. Pleska, M. V. Ricciardi, K.-H. Schmidt, and J. Ta\"{i}eb
{\it et al.},
2011 Phys. Rev. C \textbf{84} 054605.

\bibitem{Henz08}
Henzlova D, Schmidt K H, Ricciardi M V %,
%A. Keli\'{c}, V. Henzl, P. Napolitani, L. Audouin, J. Benlliure,
%A. Boudard, E. Casarejos, J. E. Ducret, T. Enqvist, A. Heinz,
%A. Junghans, B. Jurado, A. Kr\'{a}sa, T. Kurtukian, S. Leray,
%M. F. Ord\'{o}\~{n}ez, J. Pereira, R. Pleska\v{c}, F. Rejmund,
%C. Schmitt, C. St\'{e}phan, L. Tassan-Got, C. Villagrasa, C. Volant,
%A. Wagner, and O. Yordanov,
{\it et al.},
2008 Phys. Rev. C \textbf{78} 044616.




\end{thebibliography}
\end{document}